\documentclass[11pt]{article}
\usepackage{amsmath,amsfonts}
\usepackage{fullpage}
\usepackage{lscape}
\usepackage[dvips]{epsfig}

\def\##1{\underline #1}
\def\=#1{\underline{\underline #1}}
\def\.{\mbox{ \tiny{$^\bullet$} }}

\def\cos{\mbox{cos}}

\def\le{\left(}
\def\ri{\right)}
\def\les{\left[}
\def\ris{\right]}
\def\lec{\left\{}
\def\ric{\right\}}

\def\c#1{\cite{#1}}

\def\r#1{(\ref{#1})}
\begin{document}

\bigskip

\bigskip
\begin{center}
\Large{ {\bf Note on the asymptotic approximation of a double
integral with an angular spectrum representation}}
\end{center}

\begin{center}

\bigskip

Fei Wang\footnote{Fax: +1 814 865 9974 ; e--mail: fuw101@psu.edu}

{\em CATMAS --- Computational and Theoretical Materials Sciences Group \\
     Department of Engineering Science and Mechanics \\
     Pennsylvania State University, University Park, PA
     16802--6812, USA}

\end{center}

\bigskip

\noindent {\bf Abstract.} In this note, we  are concerned with the
asymptotic approximation of a class of double integrals which can
be represented as an angular spectrum superposition. These double
integrals typically appear in electromagnetic scattering
problems. Based on the synthetic manipulation
of the method of steepest descent path,
approximate expressions of the double integrals are derived in terms of the
leading term of the contribution to the asymptotic expansions.

\section{Introduction}
The methods of steepest descent path (SDP) and stationary phase
are commonly useful for the asymptotic evaluation of the
complicated integrals in electromagnetics particularly in the far
zone\c{Lang}--\c{LM}. For the one--dimensional case, it seems no
restrictions need to be enforced on the integrands to apply these
classic asymptotic methods, although sometimes the construction of
the SDP is difficult as also is the establishment of  the uniform
asymptotic expansions in the neighborhoods of the singularities of
the integrands close to the saddle points \c{Wong, Van}. But for
the asymptotic approximation of double integrals
\begin{equation}
I(\lambda)=\iint\limits_D^{} \varphi(x,y)\exp\les i\lambda\,
h(x,y) \ris \mbox{d}x\mbox{d}y\,, \quad \lambda \rightarrow \infty
\,,
\end{equation}
the stationary phase method is generally used provided that the
integrands have certain restricted forms, i.e., the phase
functions $h(x, y)$ are real--valued \c{BW, Wong, BMM, Jone1}.
When the complex--valued phase functions are concerned, asymptotic
expansions of the double integrals are still possible to be
obtained if the integrands are smooth sufficiently and vanish on
the boundary of the integration domain $D$. In this case, the
dominant contribution to the asymptotic expansions may come from
the critical points of the first kind. However, if the phase
function is complex--valued in general but not differentiable
everywhere in the domain, some additional work may be necessary to
account for the contribution of the boundary stationary points to
the asymptotic expansions of the double integrals
\c{Wong1}--\c{Jone2}. These boundary stationary points are
certainly located in the vicinity of regions wherein the phase
function is not differentiable, i.e., branch points and
singularities. Therefore a curve of stationary points may be
formed on the boundaries of the sub--domains $D_j \subset D$ which
are disjoined with each other by the ``deleted" neighborhoods of
nondifferentiation. In addition, because the phase functions now
are complex--valued, approximation of the Laplace--type integrals
may need to be manipulated on some sub--domains to accomplish the
classic asymptotic evaluation of these double integrals \c{Wong,
Jone2}\renewcommand{\thefootnote}{\fnsymbol{footnote}}
\footnote[2]{The boundary stationary points are typically the
critical points of the second kind. If there is a curve of
stationary points formed on the boundary of the domain $D$, we
know that the leading term of the contribution to the asymptotic
expansion of the double integral in $(1)$ is $O(\lambda^{-1})$ as
$\lambda\rightarrow \infty$. Similarly, for the asymptotic
expansion of the Laplace--type integral
$I(\lambda)=\iint\limits_D^{} \varphi(x,y)\exp\les -\lambda\,
h(x,y) \ris \mbox{d}x\mbox{d}y$, if the stationary points are the
critical points of the first kind, then the leading term of
$I(\lambda)$ is also $O(\lambda^{-1})$ as $\lambda \rightarrow +
\infty$. For more details, see \c{Wong, Jone2}.}.

In this note we present a simpler way to establish the asymptotic
approximation of a class of double integrals which are derived
from an angular spectrum representation of the field--related
entities. After departing from the SDP method operated in a
synthetic fashion, the approximation of these double integrals is
achieved by using the contour deformation to confine the
integration on a small domain defined by two local SDPs. The
saddle points along each SDP are employed to construct the
truncated Taylor expansion of the complex--valued phase function
which is differentiable on the mapped domain. The double integral
is then asymptotically evaluated by the lowest order in the
far--zone limit. A condition for the asymptotic approximation to
be valid is also briefly commented.

\section{Theory}
To begin with let us suppose the vector ${\bf r}=(x,y,z)$ defined
on the upper half space $\mathbb{R}_+^3=\lec z>0 \ric$ and a
double integral $G({\bf r})$ generally having the form
\begin{equation}
\label{a21} G({\bf r})=\iint\limits_{\mathbb{R}^2}^{} f(k_x, k_y)
\exp \les i (k_x x+k_y y+ k_z z)\ris\,\mbox{d}k_x\,\mbox{d}k_y\,,
\end{equation}
where $k_z=\sqrt{k_0^2-k_x^2-k_y^2}$ is defined on the top Riemann
surface such that $\mbox{Im}(k_z)\geq 0$ for $k_x^2+k_y^2\in
\mathbb{R}$.  Generally speaking, $G({\bf r})$ represents a class
of functions that have the physical importance in reducing a
field--related entity into an angular spectrum superposition. It
is reasonably assumed that $|G({\bf r})|<\infty$ is well defined
by the integrand belonging to $L^1(\mathbb{R}^2)$. In most cases,
we may also assume that $f(k_x, k_y)$ in \r{a21} is independent of
${\bf r}$ and has no singularities on the complex domain
$\Omega=\mathfrak{D} \times \mathbb{R}^c \cup \mathbb{R}^c \times
\mathfrak{D} $, where $\mathfrak{D}$ is a complex neighborhood of
the real axis, and $\mathbb{R}^c= \mathbb{C} \setminus \mathbb{R}
$.

According to Fubini's theorem, $G({\bf r})$ in \r{a21} could be
rewritten as
\begin{equation}
\label{a22} G({\bf r})=\int_{-\infty}^{\infty} \exp(ik_x x)
\,\mbox{d}k_x \int_{-\infty}^{\infty} f(k_x, k_y) \exp \les i (k_y
y+ k_z z)\ris \,\mbox{d}k_y\,.
\end{equation}
By denoting $g(k_x)=\int\limits_{-\infty}^{\infty} f(k_x, k_y)
\exp \les i (k_y y+ k_z z) \ris \mbox{d}k_y$, we know that
$g(k_x)\in L^1(\mathbb{R}) $ almost everywhere. Indeed, $g(k_x)$
may have isolated singularities which depend on $f(k_x, k_y)$.
However, because $f(k_x, k_y)$ is assumed to be nonsingular in
$\Omega$, it is clear that the isolated singularities of $g(k_x)$,
if exist, are only real--valued. Conventionally $g(k_x)$ can be
expanded in the form $g(k_x)=\sum\limits_{n=0}^\infty a_n\le k_x
-s_n \ri^{\gamma_n -1}$ in the vicinity of the singular point $s_n
 \in \mathbb{R}$. Since $g(k_x)\in L^1(\mathbb{R})$, it seems
that those singularities $\lec s_n \ric$  are not likely to be
poles, so that $\gamma_n >0$. On the other hand, we can reform
$g(k_x)$ as the multiplication of two parts
\begin{equation}
\label{a23} g(k_x)=\tilde{f}(k_x)\exp(i k_{\rho x} \rho_x)\,,
\end{equation}
where $\rho_x=\sqrt{r^2-x^2}$, $r=|{\bf r}|$ and $k_{\rho
x}=\sqrt{k_0^2-k_x^2}$ with $\mbox{Im}(k_{\rho x}) \geq 0$ for
real $k_x$. By substituting \r{a23} into \r{a22} and noting that
$\tilde{f}(k_x)$ has no poles, it is always feasible to reduce
$G({\bf r})$ in \r{a22} into a contour integral along the SDP of
the phase function. Therefore, we obtain synthetically the first
asymptotic approximation
\begin{eqnarray}
G({\bf r})&=&\int_{-\infty}^{\infty} \tilde{f}(k_x)\exp\les i( k_x
x+
k_{\rho x} \rho_x) \ris \,\mbox{d}k_x \nonumber \\
&=&\int\limits_{\mbox{SDP}_1}^{} \tilde{f}(k_{x})\exp\les ik_0
r\cos(\alpha_x-\psi_x) \ris \,\mbox{d}k_x
\nonumber \\
&\approx&\int\limits_{\mbox{SDP}_1^{loc}}^{}
\tilde{f}(k_{x})\exp\les ik_0r\cos(\alpha_x-\psi_x) \ris
\,\mbox{d}k_x\,,\label{a24}
\end{eqnarray}
as $k_0r \rightarrow \infty$. In  \r{a24}, $k_x=k_0\, \cos
\alpha_x$, $\psi_x=\cos^{-1}\le \frac{x}{r} \ri$ and
$\mbox{SDP}_1:(-\infty, \infty)\mapsto \mathbb{C}$ is determined
by the parametrization of $\alpha_x$ (or $k_x$) through the
equation $\mbox{Re}\les \cos(\alpha_x-\psi_x)-1 \ris=0$.
$\mbox{SDP}_1^{loc}$ denotes the local path of the $\mbox{SDP}_1$
that passes through the saddle point $k_{xs}=k_0\,\cos\psi_x$.

On account of  \r{a22}--\r{a24}, by using Fubini's theorem again
we obtain
\begin{eqnarray}
G({\bf r})&\approx&\int_{-\infty}^{\infty} \exp(ik_y y)
\,\mbox{d}k_y \int\limits_{\mbox{SDP}_1^{loc}}^{} f(k_{x}, k_y)
\exp \les i
(k_x x + k_z z) \ris \,\mbox{d}k_x \nonumber\\
&=&\int_{-\infty}^{\infty} \tilde{\tilde{f}}(k_y) \exp\les i (k_y
y+k_{\rho y} \rho_y) \ris \,\mbox{d}k_y\,,\label{a25}
\end{eqnarray}
where the function $\tilde{\tilde{f}}(k_y)$ is defined by the
equation
\begin{equation}
\int\limits_{\mbox{SDP}_1^{loc}}^{} f(k_{x}, k_y) \exp \les i (k_x
x + k_z z) \ris \,\mbox{d}k_x=\tilde{\tilde{f}}(k_y)\exp(i k_{\rho
y} \rho_y)\,. \label{a26}
\end{equation}
Here $\rho_y=\sqrt{r^2-y^2}$ and $k_{\rho y}=\sqrt{k_0^2-k_y^2}$
with $\mbox{Im}(k_{\rho y}) \geq 0$ for real $k_y$. Actually the
definition of $\tilde{\tilde{f}}(k_y)$ in \r{a26} is very similar
to that of $\tilde{f}(k_{x})$ in \r{a23}. Following the same
approach as illustrated in \r{a22}--\r{a24}, we can obtain the
second asymptotic approximation of $G({\bf r})$  as follows:
\begin{eqnarray}
G({\bf r})&\approx& \int\limits_{\mbox{SDP}_2}^{}
\tilde{\tilde{f}}(k_{y}) \exp\les ik_0r\cos(\alpha_y-\psi_y) \ris
\,\mbox{d}k_y \nonumber\\
 &\approx& \int\limits_{\mbox{SDP}_2^{loc}}^{}
\tilde{\tilde{f}}(k_{y}) \exp\les ik_0r\cos(\alpha_y-\psi_y) \ris
\,\mbox{d}k_y\,. \label{a27}
\end{eqnarray}
In  \r{a27}, $k_y=k_0\, \cos \alpha_y$, $\psi_y=\cos^{-1} \le
\frac{y}{r} \ri$ and $\mbox{SDP}_2:(-\infty, \infty)\mapsto
\mathbb{C}$ is determined by the parametrization of $\alpha_y$ (or
$k_y$) through the equation $\mbox{Re}\les \cos(\alpha_y-\psi_y)-1
\ris=0$. $\mbox{SDP}_2^{loc}$ denotes the local path of the
$\mbox{SDP}_2$ that passes through the saddle point
$k_{ys}=k_0\,\cos\psi_y$.

It seems that $G({\bf r})$ can be asymptotically approximated by
either \r{a24} or \r{a27}. Clearly, the asymptotic expression of
$G({\bf r})$ is strongly dependent on the properties of
$\tilde{f}(k_x)$ and $\tilde{\tilde{f}}(k_y)$ in the fact that the
leading contribution to $G({\bf r})$ would be significantly
influenced or even dominated by the singularities of
$\tilde{f}(k_x)$  (or $\tilde{\tilde{f}}(k_y)$) that are close to
the saddle point $k_{xs}$ (or $k_{ys}$) \c{Wong, Wong2}. On the
other hand, the synthetic equations \r{a23} and \r{a26} imply that
both $\tilde{f}(k_x)$ and $\tilde{\tilde{f}}(k_y)$ have been
modulated by the larger parameter $k_0r$. Therefore, the
asymptotic expansion of $G(r)$ cannot be obtained in closed--form,
unless the asymptotic expansion of $\tilde{f}(k_x)$ (or
$\tilde{\tilde{f}}(k_y)$ ) is achieved beforehand.

However, the asymptotic approximation of $G({\bf r})$  can still
be accomplished by restoring $G({\bf r})$ with a double integral
representation. In fact, due to \r{a25}--\r{a27}, the
approximation of $G({\bf r})$ is now reduced to a double integral
\begin{equation}
G({\bf r})\approx \iint\limits_{D^\prime}^{} f(k_{x}, k_{y}) \exp
\les i (k_x x+k_y y+ k_z z)\ris\,\mbox{d}k_x\,\mbox{d}k_y\,,
\label{a28}
\end{equation}
where the domain $ \mathbb{C} \supset
D^\prime:=\mbox{SDP}_1^{loc}\times\mbox{SDP}_2^{loc}$ is located
in a neighborhood of the critical point $(k_{xs}, k_{ys})$. The
local pathes of the SDPs can be represented parametrically by the
real--valued $\xi$ and $\eta$ as
\begin{eqnarray}
k_x(\xi)&=&k_{xs}+k_{zs}(1-i)\xi\,,\label{a29}\\
k_y(\eta)&=&k_{ys}+k_{zs}(1-i)\eta\,, \label{a210}
\end{eqnarray}
where $k_{zs}=\sqrt{k_0^2-k_{xs}^2-k_{ys}^2}$  \c{Felsen1}.
Therefore $D^\prime$ can be viewed a continuous map from the local
domain $\xi \times \eta\subset \mathbb{R}^2$ by means of \r{a29}
and \r{a210}.

Let $U(\xi, \eta)=(k_0 r)^{-1}(k_x x+k_y y+ k_z z)-1$ be defined
on $D^\prime$; then $G({\bf r})$ in \r{a28} can be written in the
form
\begin{eqnarray}
G({\bf r})&\approx& \exp(ik_0 r) \iint\limits_{D^\prime}^{}
f(k_{x}, k_{y}) \exp \les i k_0 r\, U(\xi, \eta)
\ris\,\mbox{d}\xi\,\mbox{d}\eta\,. \nonumber\\
 &\approx& f(k_{xs}, k_{ys})\exp(ik_0 r)
\iint\limits_{D^\prime}^{} \exp \les i k_0 r\, U(\xi, \eta)
\ris\,\mbox{d}\xi\,\mbox{d}\eta\,,\label{a211}
\end{eqnarray}
by utilizing the approximation $f(k_{x}, k_{y})\approx f(k_{xs},
k_{ys})$ on $D^\prime$. Indeed, to be more accurate, we need to
express $f(k_x, k_y)$ in term of the Taylor expansion by assuming
$f(k_x, k_y)$ is analytic in a neighborhood of the critical point.
There will be no additional difficulty in working out the
asymptotic approximation of $G({\bf r})$ in \r{a211} when
$f(k_{x}, k_{y})$ can be represented by its Taylor expansion.
Furthermore, if $U(\xi, \eta)$ or $k_z=\sqrt{k_0^2-k_x^2-k_y^2}$
is differentiable on the domain $D^\prime$, then we can
approximate $U(\xi, \eta)$ by its truncated Taylor expansion on
the critical point $(\xi, \eta)=(0, 0)$. In fact,
$k_z=\sqrt{k_z^2}$ is holomorphic on the complex domain
$\mathbb{C} \supset D_2:=\lec k_z^2\ric$ continuously mapped from
$D^\prime$ as shown in Figure 1, simply because $D_2$ is located
on a Reimann surface that neither crosses the branch cut nor
encloses the branch point $(0, 0)$. Therefore $U(\xi, \eta)$ is
analytic on $D^\prime$ in accordance with the chain rule of
differentiation. Then $U(\xi, \eta)$ is approximated by the Taylor
expansion at $(0, 0)$ truncated to the second order as:
\begin{equation}
U(\xi, \eta)\approx U_{T}(\xi,
\eta)=i(a\,\xi^2+b\,\eta^2-2c\,\xi\eta) \,,\quad  (\xi, \eta) \in
D^\prime \,,\label{a212}
\end{equation}
by recalling $\frac{\partial U}{\partial \xi}=\frac{\partial
U}{\partial \eta}=0$ at $(0, 0)$. Here
\begin{equation}
a=1-\frac{y^2}{r^2}\,, \quad b=1-\frac{x^2}{r^2}\,,\quad
c=\frac{x\,y}{r^2}\,,\label{a213}
\end{equation}
are real--valued constants.

After substituting \r{a212} into \r{a211}, we then derive the
asymptotic approximation ( $k_0r \rightarrow \infty$)
\begin{eqnarray}
G({\bf r})&\approx& f(k_{xs}, k_{ys})\exp(ik_0 r)
\iint\limits_{D^\prime}^{} \exp \les - k_0 r
(a\,\xi^2+b\,\eta^2-2c\,\xi\eta) \ris\,\mbox{d}\xi\,\mbox{d}\eta \nonumber\\
&\approx&f(k_{xs}, k_{ys})\exp(ik_0 r)
\int\limits_{-\infty}^{\infty} \int\limits_{-\infty}^{\infty} \exp
\les - k_0 r (a\,\xi^2+b\,\eta^2-2c\,\xi\eta)
\ris\,\mbox{d}\xi\,\mbox{d}\eta\,,\label{a214}
\end{eqnarray}
which is the leading term of the factor $(k_0r)^{-1}$ of the
contribution to the asymptotic expansion of $G({\bf r})$. In fact,
the integral on the right side of \r{a214} can be easily
calculated by eliminating the corss--product term $\xi\eta$
through a linear transformation.

It is worthy to note that the approximation of $U(\xi, \eta)$ by
its truncated Taylor expansion \r{a212} can be reasonably achieved
only under the condition that $\theta=\frac{z}{r}>0$ is not too
small. In fact the Taylor expansion of $U(\xi, \eta)$ at (0,0)
takes the form  $U(\xi, \eta)=\sum\limits_{|\beta|=2}^{\infty}
u_{\beta} \,\#\xi^{|\beta|=2} \les 1+P_{\beta}(\#\xi) \ris$ where
$\#\xi^{\beta}=(\xi^{\beta_1}, \eta^{\beta_2})$ and
$\beta=(\beta_1, \beta_2)$ is a multiindex such that
$|\beta|=\beta_1+ \beta_2$. If the coefficient $u_{\beta}$ is kept
as a factor of $O(1)$, then the function $P_{\beta}(\#\xi)\sim
O(|\#\xi|/\theta)^{|\beta|-2}$ could be achieved at some $\beta$
for any $|\beta|\geq 2$. After considering the approximation of
\r{a214}, $|\#\xi|^2 \sim O(1/k_0{r})\in D^\prime$ is at least
guaranteed. Therefore, $\theta
>  O\le\sqrt{\frac{1}{k_0r}}\,\,\ri$ is needed for an
appropriate approximation of $U(\xi, \eta)$ by \r{a212} on the
domain $D^\prime$. In other words, the asymptotic approximation of
$G({\bf r})$ by \r{a214} becomes reasonable only when $\theta
>\theta_0$ is necessarily satisfied, where
$\theta_0=\sqrt{\frac{1}{k_0r}}$.

\section{Conclusion}
In this note, we are concerned with the asymptotic approximation
of a class of double integrals which can be represented as an
angular spectrum superposition. These double integrals are
typically useful for the calculation of field--related problems in
physics and electromagnetics. A simpler way is found to make the
asymptotic approximation feasible and reasonable by applying the
steepest descent path method on the double integral in a synthetic
fashion. After reduction to a local integration on a small domain
in which the complex--valued phase function is differentiable, the
asymptotic approximation of the double integral is finally derived
in terms of the leading term of the contribution to the asymptotic
expansion.  The validity of the asymptotic approximation is only
under the condition that $ \frac{z}{r}>\theta_0  $ is satisfied,
where $\theta_0=\sqrt{\frac{1}{k_0r}}>0$.

\newpage
\begin{center}
{\bf Figure Captions}
\end{center}
 \vskip 0.3 cm

{\bf Figure 1.}  Schematic of the continuous map $k_z^2:D^\prime
(\xi, \eta)\mapsto D_2(k_z^2)$.

 \vskip 0.3 cm

\newpage
\begin{figure}[!ht]

\centering \psfull \epsfig{file=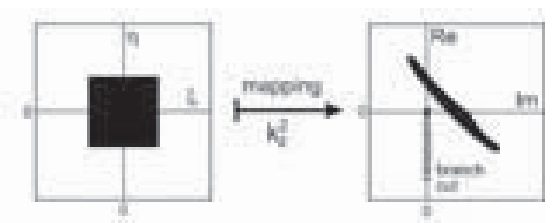,width=3in}
\end{figure}
\bigskip\bigskip
\noindent Figure 1: Schematic of the continuous map
$k_z^2:D^\prime (\xi, \eta)\mapsto  D_2(k_z^2)$.


\bigskip

\begin{thebibliography}{03}

\bibitem{Lang}
Jordan A K  and Lang R H 1979 {\em Radio Sci.\/} {\bf 14} 1077

\bibitem{BW}
Born M and Wolf E 1987 {\em Principles of Optics\/} (UK:Pergamon)

\bibitem{Lakhtakia}
Lakhtakia A 1993 {\em Arch. Elektr. \"Ubertr.\/} {\bf 47}
1


\bibitem{LM}
Lakhtakia A and McCall M W 2003 {\em Arch. Elektr. \"Ubertr.\/} {\bf 57}
23


\bibitem{Wong}
Wong R 2001 {\em Asymptotic Approximations of Integrals\/} (SIAM
Philadelphia, PA)

\bibitem{Van}
van der Waerden B L 1951 {\em Appl. Sci. Res.\/} {\bf B2} 33


\bibitem{BMM}
Bouche D, Molinet F and Mittra R 1994 {\em Asymptotic Methods in
Electromagnetics\/} (Springer, Heidelberg, Germany)

\bibitem{Jone1}
Jones D S and Kline M 1958 {\em J. Math. Phys.\/} {\bf 37} 1

\bibitem{Wong1}
McClure J P and Wong R 1991 {\em SIAM J. Math. Anal.\/} {\bf 22}
500

\bibitem{Erdelyi}
Erd$\acute{e}$lyi A 1959 {\em Proceedings of the Fourth Canadian
Mathematical Congress\/} (Univeristy of Toronto Press, Toronto) pp
137-146

\bibitem{Jone2}
Jones D S 1982 {\em The Theory of Generalized Functions \/},
(Cambridge University Press, Cambridge, UK)

\bibitem{Wong2}
Wong R and Wyman M 1972 {\em Can. J. Math.\/} {\bf 24} 185

\bibitem{Felsen1}
Capolino F, Maci S and Felsen L B 2000 {\em Radio Sci.\/} {\bf 35}
579



\end{thebibliography}
\end{document}